\documentclass[useAMS,usenatbib]{mn2e}
\usepackage{amssymb}
\usepackage{graphicx}
\usepackage{graphics}
\usepackage{epsfig}

\title[Particle Swarm Optimization]{Particle Swarm
Optimization: An efficient method for tracing periodic orbits in
3D galactic potentials}
\author[Ch.~Skokos et al.]{Ch.~Skokos,$^{1,2}$\thanks{E-mail:
hskokos@cc.uoa.gr}\thanks{Corresponding author}
K.~E.~Parsopoulos,$^{3}$ P.~A.~Patsis$^{1}$ and
M.~N.~Vrahatis$^{3}$\\
$^{1}$Research Center for Astronomy, Academy of Athens, Soranou
Efesiou 4,
 GR-11527 Athens, Greece\\
$^{2}$Department of Mathematics, Division of Applied Analysis and
Center for Research and Applications of Nonlinear Systems
(CRANS),\\
University of Patras, GR-26500 Patras, Greece\\
$^{3}$Department of Mathematics,  University of Patras Artificial
Intelligence Research Center (UPAIRC),\\ University of Patras,
GR-26110 Patras, Greece}
\begin{document}

\date{Accepted; Received; in original form }

\pagerange{\pageref{firstpage}--\pageref{lastpage}} \pubyear{2004}

\maketitle

\label{firstpage}

\begin{abstract}
We propose the Particle Swarm Optimization (PSO) as an alternative
method for locating periodic orbits in a three--dimensional (3D)
model of barred galaxies. We develop an appropriate scheme that
transforms the problem of finding periodic orbits into the problem
of detecting  global minimizers of a function, which is defined on
the Poincar\'{e} Surface of Section (PSS) of the Hamiltonian
system. By combining the PSO method with deflection techniques, we
succeeded in tracing systematically several periodic orbits of the
system. The method succeeded in tracing the initial conditions of
periodic orbits in cases where Newton  iterative techniques had
difficulties. In particular, we found families of 2D and 3D
periodic orbits associated with the inner 8:1 to 12:1 resonances,
between the radial 4:1 and corotation resonances of our 3D Ferrers
bar model. The main advantages of the proposed algorithm is its
simplicity, its ability to work using function values solely, as
well as its ability to locate many periodic orbits per run at a
given Jacobian constant.
\end{abstract}


\section{Introduction}
\label{intro}

The periodic orbits of an autonomous Hamiltonian system, as well
as their stability play a crucial role for the dynamical behavior
of the system. Orbits that are located near stable periodic orbits
are ordered, while, near unstable periodic orbits chaotic motion
occurs. Therefore, by locating the main periodic orbits of a
system and  following their stability properties as one of its
parameters changes, we obtain valuable information on the ordered
or chaotic nature of motion in the system.

The orbital study of  barred potentials has provided useful
information on the structure of galactic bars
\citep[c.f.][]{C80,A84,CG89,SW93,P96,P04}. In 2--dimensional (2D)
models, the galactic bar is supported by regular orbits trapped
around the so called `{\tt x1}' periodic orbits, which are
elongated along the bar major axis \citep{CG89}. Based on the fact
that these orbits do not extend beyond the corotation resonance,
\citet{C80} predicted that bars should end at, or before,
corotation. In 3D models, the planar {\tt x1} family has in
general large unstable parts and, thus, its orbits are not
sufficient in building the bar. However, other families of
periodic orbits that bifurcate from {\tt x1} have large stable
parts and support the bar. These families build the so--called
`{\tt x1}--tree' (Skokos, Patsis \& Athanassoula 2002a,b).
Specific families of the  {\tt x1}--tree are associated with
certain morphological features observed in real galaxies (Patsis,
Skokos \& Athanassoula 2002, 2003a). Although the basic
morphological features of barred galaxies are related to the
presence of orbits of the {\tt x1}--tree, orbits not belonging to
this tree can also influence the galaxy's morphology. Recently, it
has been shown (Patsis, Skokos \& Athanassoula 2003b) that inner
rings in barred galaxies are associated with specific families of
periodic orbits influenced by the 4:1, 6:1 and 8:1 resonances,
which are located just beyond the end of the bar (for a definition
of the resonances see e.g. \citet{C02} Section 3.1.1).

The basic families belonging to the {\tt x1}--tree are usually
located very easily, but, as we approach corotation, tracing
periodic orbits that are influenced by high order resonances
becomes a difficult and challenging problem. The difficulty in
finding such orbits is mainly due to the fact that at this region
there exist many periodic orbits close to each other, most of
which are unstable. Thus, one is looking for small or even tiny
islands of stability in a region of the phase space of the system,
which is mainly characterized by chaotic behavior. Usually once a
periodic orbit is located, the whole family in which it belongs
can be found as a parameter of the system (e.g., the Jacobi
integral) changes. In this way, we can follow the morphological
evolution and the stability transitions of a family and determine
its physical importance for the system.

The problem of finding the initial conditions of  periodic orbits
for a given parameter set of a Hamiltonian system is the starting
point of the present paper. Using an appropriate scheme we
transform the aforementioned problem into a minimization problem,
where the minimizers of a particular function defined on the
Poincar\'{e} Surface of Section (PSS) of the system correspond to
the periodic orbits of the Hamiltonian. Then, by applying an
efficient optimization method,  Particle Swarm Optimization (PSO),
and using deflection techniques, we locate periodic orbits, and
follow them as a parameter of the system varies. This procedure is
applied successfully on a 3D galactic potential of a Ferrers bar.

The paper is organized as follows. In Sec. \ref{pot} we describe
the particular model that is used for our orbital calculations.
Sec. \ref{alg} is devoted to the thorough description of the
proposed algorithm. In particular, in Sec. \ref{trans} the
procedure of transforming the problem of detecting periodic orbits
into a minimization problem is described, in Sec. \ref{PSO} the
PSO method for addressing it is presented, while  Sec. \ref{def},
is devoted to the deflection technique, which allows us to detect
further periodic orbits. In Sec. \ref{res}, we report the obtained
periodic orbits and discuss their physical importance. Finally, in
Sec. \ref{con}, we discuss the effectiveness of the proposed
numerical scheme and present our conclusions.

\section{The galactic potential}
\label{pot}

The  3D barred galaxy model that we use is  described in detail in
\citet{SPA02a}. It consists of a Miyamoto disk, a Plummer bulge
and a Ferrers bar. The potential of the Miyamoto disk
\citep{mina75} is given by the formula,

\begin{equation}
\label{potd}
V_{D}=-\frac{GM_{D}}{\sqrt{x^{2}+y^{2}+(A+\sqrt{B^{2}+z^{2}})^{2}}},
\end{equation}
where \( M_{D} \) represents the total mass of the disk, $A$ and
$B$ are scale lengths such that the ratio $B/A$ gives a measure of
the flatness of the model, and $G$ is the gravitational constant.
The bulge is a Plummer sphere, i.e., its potential is given by,
\begin{equation}
\label{pots} V_{S}=-\frac{GM_{S}}{\sqrt{x^{2}+y^{2}+z^{2}+\epsilon
_{s}^{2}}},
\end{equation}
where \( \epsilon _{s} \) is the bulge scale length and \( M_{S}
\) is its total mass.  Finally, the bar is a triaxial Ferrers bar
with density,
\begin{equation}
\label{densd} \rho (m)=\left\{ \begin{array}{lcc}
\displaystyle{\frac{105M_{B}}{32\pi abc}(1-m^{2})^{2}}, \quad  &
{\mbox for} &
m \leq 1,\\
 & & \\
\displaystyle{0}, & {\mbox for}  & m>1,
\end{array}\right.
\end{equation}
where,
\begin{equation}
\label{semiaxis}
m^{2}=\frac{y^{2}}{a^{2}}+\frac{x^{2}}{b^{2}}+\frac{z^{2}}{c^{2}}\,
\, ,\, \,\quad
 \,
 a>b>c.
\label{eq:m}
\end{equation}
In Eq.~(\ref{eq:m}), \( a \), \( b \) and \( c \) are the
principal semi-axes, while \( M_{B} \) denotes the mass of the bar
component. The corresponding potential, \( V_{B} \), as well as
the forces are given in \citet{pf84} in a closed form, which is
well suited for numerical treatment.

Regarding the Miyamoto disk, we use $A=3$ and $B=1$, and for the
axes of the Ferrers bar we set $a=6$, $b=1.5$ and $c = 0.6$. The
masses of the three components satisfy \( G(M_{D}+M_{S}+M_{B})=1
\). In particular, we have $GM_{D}=0.82$, $GM_{S}=0.08$,
$GM_{B}=0.10$ and $\epsilon _{s}=0.4$.  The length unit is taken
as 1~kpc, the time unit as 1~Myr and the mass unit as $ 2\times
10^{11} M_{\odot}$. The bar rotates with a pattern speed,
$\Omega_{b}$=0.054, around the $z$-axis, which corresponds to
54~km~sec$^{-1}$~kpc$^{-1}$, and places corotation at 6.13~kpc.
The Hamiltonian governing the motion of a test particle can be
written in the form,
\begin{equation}
H=\frac{1}{2} ( p_x^2+ p_y^2+p_z^2 ) + V_D + V_S + V_B -\Omega_b
(x p_y -y p_x), \label{eq:Ham}
\end{equation}
with $p_x=\dot{x}-\Omega_b y$, $p_y=\dot{y}+\Omega_b x$ and
$p_z=\dot{z}$ being the canonical momenta. The numerical value of
$H$ will be vaguely reported as the `energy', $E_j$, of the
system.

\section{Description of the algorithm}
\label{alg}

{\em Swarm Intelligence\/} methods are stochastic optimization,
machine learning and classification procedures that model
intelligent behavior (Bonabeau, Dorigo \& Theraulaz 1999; Kennedy
\& Eberhart 2001). They are closely related to the methods of {\em
Evolutionary Computation}, which consists of algorithms motivated
from biological genetics and natural selection. A common
characteristic of all these algorithms is the exploitation of a
population of search points that probe the search space
simultaneously \citep{Schwefel94,Fogel00}.

Particle Swarm Optimization (PSO) belongs to the category of Swarm
Intelligence methods. The development of PSO sprang from   the
simulation of social dynamics of flocking organisms, such as
insect swarms, which are governed by fundamental rules like
nearest--neighbor velocity matching. In nature, information is
communicated among the members of bird flocks and fish schools,
enhancing their ability to search for food and enabling them to
move synchronized without colliding \citep{Millonas94}.  The
social behavior of animals, and in some cases of humans, is
governed by similar rules. There is a general belief, and numerous
examples coming from nature enforce it, that social sharing of
information among the individuals of a population, provide an
evolutionary advantage.

The dynamics of  population in PSO resembles the collective
behavior and self--organization of socially intelligent organisms.
The individuals of the population exchange information and benefit
from their discoveries, as well as the discoveries of other
companions, while exploring promising areas of the search space.
In our case, the Poincar\'{e} section is our search space, and
periodic orbits of a Hamiltonian system are computed through the
minimization of a function. In the context of function
minimization, promising areas of the search space are
characterized by  low function values.

\subsection{The objective function}
\label{trans}

A suitable way to study the stability of orbits in the 6D phase
space of the Hamiltonian system (\ref{eq:Ham}) is the well--known
method of the Poincar\'{e} Surface of Section
\citep[e.g.][]{LL92}. Instead of following the time evolution of
an orbit in the whole phase space, we confine our study on an
appropriately chosen subspace of it. In our case, the PSS is the
subspace  $(x,z,\dot{x},\dot{z})$ of $\mathbb{R}^6$, defined by
the conditions, $y=0$, $\dot{y}>0$. The major axis of the bar lies
along the $y$ axis. Thus, for a given value of the energy $E_j$,
an orbit with initial conditions,
$X_0=(x_0,z_0,\dot{x}_0,\dot{z}_0)^{\top}$ (where $^{\top}$
denotes the transpose of a matrix), on the PSS is fully defined as
$y=0$ and $\dot{y}$ can be obtained by solving Eq. (\ref{eq:Ham})
keeping only the positive found value of $\dot{y}$.

In this way, only the 4 initial conditions of an orbit on the PSS
are necessary for identifying the orbit. Then, the time evolution
of the orbit is derived by solving the Hamilton's equations of
motion. The next intersection of the orbit with the PSS,  $y=0$,
$\dot{y}>0$, is denoted as,
\begin{equation}
\Phi(X_0)=\big(\Phi_x(X_0),\Phi_z(X_0),\Phi_{\dot{x}}(X_0),\Phi_{\dot{z}}(X_0)\big)^{\top}:
\mathbb{R}^4 \rightarrow \mathbb{R}^4, \label{eq:Phi}
\end{equation}
which is obviously a point belonging to the 4--dimensional PSS of
the system. By using the notation $\Theta_4 = (0,0,0,0)^{\top}$,
the initial conditions, $X$, of a $p$--periodic orbit of the
system satisfy the equations:
\begin{eqnarray}
\Phi^p(X) & = & X \Rightarrow \Phi^p(X) - X = \Theta_4 \Rightarrow  \nonumber\\
& & \left(
\begin{array}{c}
\Phi^p_x(X)\\
\Phi^p_z(X)\\
\Phi^p_{\dot{x}}(X)\\
\Phi^p_{\dot{z}}(X)
\end{array} \right) -
\left(
\begin{array}{c}
x\\
z\\
\dot{x}\\
\dot{z}
\end{array} \right) =
\left(
\begin{array}{c}
0\\
0\\
0\\
0
\end{array} \right).
\label{eq:eq}
\end{eqnarray}
Thus, finding the initial conditions, $X$, of a $p$--periodic
orbit is equivalent to solving Eq.\,(\ref{eq:eq}), which in turn
is equivalent to computing the global minimizers of the function
\citep{PV02b},
\begin{eqnarray}
f(X)&= &\left( \Phi^p_x(X)-x\right)^2 + \left(
\Phi^p_z(X)-z\right)^2 +
\nonumber\\
& &\left( \Phi^p_{\dot{x}}(X)-\dot{x}\right)^2 +  \left(
\Phi^p_{\dot{z}}(X)-\dot{z}\right)^2. \label{eq:function}
\end{eqnarray}
This function is called the objective function, and it is actually
the square of the Euclidean distance on the PSS between the
initial point of an orbit and its $p$th intersection with the PSS.
Obviously, if the studied orbit is $p$--periodic this distance is
zero. Thus, the aforementioned technique transforms the problem of
finding periodic orbits into the problem of computing  the global
minimizers of the function~$f(X)$.

\subsection{The PSO method}
\label{PSO}

Let us now describe the procedure of finding a minimizer of the
 objective function $f(X)$  defined on a
4--dimensional search space, $S \subset \mathbb{R}^4$, which is a
subspace of the PSS. As already mentioned, PSO is a population
based method, i.e., it exploits a population of individuals to
probe for promising regions of the search space, simultaneously.
The population is called a {\em swarm\/} and the individuals
(i.e., the search points) are called {\em particles\/}. In our
case the swarm is a set of initial conditions on the PSS and a
particle is a point $X=(x,z,\dot{x},\dot{z})^{\top}$ on the PSS,
which corresponds to an initial condition of an orbit of
Hamiltonian (\ref{eq:Ham}). In every iteration of the PSO method,
we check whether any of the particles is  a minimizer of the
objective function, $f(X)$, of Eq.~(\ref{eq:function}). If it is,
the minimizer is recorded, otherwise,  each particle is moved to a
new position in the search space $S$ using an adaptable
displacement called {\em velocity}, retaining also a memory of the
best position it ever encountered. In our case the best positions
possess lower function values.

The particles of the swarm exchange information among them. There
are two variants of the method with respect to the number of
particles that share information: the {\em global\/}  and the {\em
local\/} variant \citep{KE01}. In the global variant of PSO, the
best position ever attained by all particles of the swarm is
communicated among them. In the local variant, each particle is
assigned to a neighborhood consisting of a prespecified number of
particles and the best position ever attained by the particles
that comprise the neighborhood is communicated among them. In the
latter case, the information attained by the particles is spread
in the swarm slowly, maintaining high diversity of the particles
for more iterations of the algorithm than in the global variant,
which converges faster to the best position. Consequently, the
local variant has better exploration abilities while the global
variant has better convergence rates (exploitation). In the
present paper we use the local variant of  PSO as it is more
appropriate for the location of periodic orbits in the chaotic
region near the corotation of the bar potential.  In this region
there are many periodic orbits, very close to each other, most of
which are unstable. Thus, in order to locate such orbits, we need
an algorithm that performs thorough exploration of the phase space
with the cost of slightly slower convergence.

Let us consider a swarm consisting of $N$ particles. Each particle
is in effect a 4--dimensional vector,
\begin{equation}
X_i = (x_{i1}, x_{i2}, x_{i3}, x_{i4})^{\top}=
(x_i,z_i,\dot{x}_i,\dot{z}_i)^{\top} \in S, \quad i=1,\ldots,N.
\label{eq:x}
\end{equation}
The velocities of the particles are also 4--dimensional vectors,
\begin{equation}
V_i = (v_{i1}, v_{i2},v_{i3},v_{i4})^{\top}, \quad i=1,\ldots,N.
\label{eq:v}
\end{equation}
The best previous position encountered by the $i$--th particle is
a point in $S$, denoted by
\begin{equation}
P_i = (p_{i1}, p_{i2}, p_{i3}, p_{i4})^{\top} \in S.
\end{equation}

The positions, $X_i$, and the velocities, $V_i$, of the particles
are randomly initialized, following a uniform distribution within
the search space. The best positions, $P_i$, are initially set
equal to $X_i$. Each particle is evaluated according to the
objective function $f(X)$ of Eq.~(\ref{eq:function}), i.e., the
value $f(X_i)$ is computed for all particles. Obviously, at the
initialization phase it holds that $f(P_i) = f(X_i)$.

Let $\mathcal{N}_i = \{ X_{i-r}, \ldots, X_{i-1}, X_i, X_{i+1},
\ldots, X_{i+r}\}$, be a neighborhood of radius $r$ of the $i$th
particle, $X_i$ (local variant). Then, $g_i$ is defined as the
index of the best particle in the neighborhood of $X_i$, i.e.,
\begin{equation}
f(P_{g_i}) \leqslant f(P_j), \quad j=i-r,\ldots, i+r. \label{eq:f}
\end{equation}
The neighborhood's topology is usually cyclic, i.e., the first
particle $X_1$ is assumed to follow after the last particle,
$X_N$. In the general case we face in the present paper, the
search space is 4--dimensional. In this case we use a swarm
consisting of $N=20$ particles, while the neighborhood of every
particle has radius, $r=3$. For example, the neighborhood of the
particle $X_2$ consists of the particles, $X_{19}$, $X_{20}$,
$X_1$, $X_2$, $X_3$, $X_4$ and $X_5$. The specific neighborhood
size was selected in order to benefit from the exploitation
ability of the local PSO variant, while avoiding very slow
convergence rates implied by smaller neighborhoods.

If there exist a particle $X_j$, such that $f(X_j)=0$, then the
initial conditions of a $p$--periodic orbit are found. In the
opposite case, we proceed to the next iteration of the algorithm,
which is to move the particles to new positions, taking into
account their history so far. This is done according to the
equations \citep{CK02,PPMV01,PV02b}
\begin{eqnarray}
V_{i}^{(q+1)} &=& \chi \, \bigg( V_{i}^{(q)} + c_1 \, r_1
\Big(P_{i}^{(q)} - X_{i}^{(q)}\Big) + \nonumber \\
 & &
                  \ c_2 \, r_2 \Big(P_{g_{i}}^{(q)} - X_{i}^{(q)}\Big)\bigg), \label{eq:PSO1con}\\
 X_{i}^{(q+1)} &=& X_{i}^{(q)} + V_{i}^{(q+1)}, \label{eq:PSO2con}
\end{eqnarray}
where $i = 1, 2, \ldots, N$; $\chi$ is a parameter called {\em
constriction factor\/}; $c_1$ and $c_2$ are two fixed, positive
parameters called {\em cognitive\/} and {\em social\/} parameter
respectively; $r_1$, $r_2$, are random vectors with components
uniformly distributed in the interval $[0,1]$; and $q$ indicates
iterations. All vector operations are performed componentwise. The
objective function $f(X)$ is computed again at the new positions,
$X_i^{(q+1)}$, of the particles, and the best positions, $P_i$,
$i=1,\ldots,N$, are updated as follows
\[
P_i^{(q+1)} = \left\{
\begin{array}{ll}
X_i^{(q+1)}, \quad & \textrm{if} \ f\left(X_i^{(q+1)}\right) < f\left(P_i^{(q)}\right),\\
P_i^{(q)}, \quad & \textrm{otherwise}.
\end{array}
\right.
\]
Then, the new indices, $g_i$, are determined and Eqs.~
(\ref{eq:PSO1con}) and (\ref{eq:PSO2con}) are applied again and so
on. We note that in the case that the velocity $V_i$ of a particle
would move it outside the search space $S$, the particle is
actually moved only up to the border of $S$. The algorithm
terminates when a user defined criterion is achieved, which means
that a global minimizer is detected. Since in our case the global
minimum of the objective function $f(X)$  is known a priori to be
equal to zero, we consider that a minimum of $f(X)$ is located at
$X^*$, if $f(X^*) \leqslant 10^{-10}$. If this criterion is not
fulfilled, the algorithm stops as soon as a predefined maximum
number of iterations, $q_{\max}$, is reached. In this case, the
PSO method applied for the particular initial distribution of
particles is considered to have failed to compute a $p$--periodic
orbit with the desired accuracy. In our experiments, $q_{\max}$
was set equal to $500$.

Let us now discuss the role of the various parameters that appear
in Eqs.\,(\ref{eq:PSO1con}) and (\ref{eq:PSO2con}). In early
versions of PSO \citep{EK95} there was no actual mechanism to
control the magnitude of the particles' velocities. Thus, they
could take arbitrarily high values (swarm explosion), resulting in
divergence of the swarm. For this purpose, a positive parameter,
$V_{\max}$, was employed as a threshold on the absolute value of
the velocity's vector components,
\[
v_{ij} = \left\{
\begin{array}{rl}
v_{ij}, \quad       & \textrm{if}\ \ |v_{ij}| \leqslant V_{\max},\\
-V_{\max},\quad & \textrm{if}\ \ v_{ij} < -V_{\max},\\
V_{\max}, \quad  & \textrm{if}\ \ v_{ij} > V_{\max},
\end{array}
\right.
\]
for $i = 1,\ldots, N$,   $j=1,\ldots, n$, and $n$ the dimension of
the search space. In more recent versions of PSO
\citep{CK02,Trelea03}, the constriction factor $\chi$ has been
introduced as a mechanism for constraining the magnitude of the
velocities. Stability analysis of the algorithm described by
Eqs.\,(\ref{eq:PSO1con}) and (\ref{eq:PSO2con}) results in the
following formula for the determination of $\chi$ \citep{CK02},
\begin{equation} \label{eq:chi}
\chi = \frac{2 \kappa}{|2 - \phi - \sqrt{\phi^2 - 4\phi}|},
\end{equation}
for $\phi > 4$, where $\phi = c_1 + c_2$, and $\kappa = 1$. A
complete theoretical analysis of the derivation of
Eq.~(\ref{eq:chi}), can be found in \citet{CK02} and
\citet{Trelea03}. The cognitive parameter, $c_1$, determines the
effect of the distance between the current position of the
particle and its best previous position, $P_i$, on its velocity.
On the other hand, the social parameter $c_2$ plays a similar role
but it concerns the best previous position, $P_{g_i}$, attained by
any particle in the neighborhood. The particular values of the
parameters that were used in  our computations are $\chi=0.729$
and $c_1=c_2=2.05$. These values are considered optimal default
values \citep{CK02,Trelea03}.

The parameters $r_1$ and $r_2$ introduce stochasticity to the
algorithm. Therefore, if the best positions are very close to each
other and lie in the basin of attraction of a local minimizer of
the objective function, then $r_1$, $r_2$ hinder the particles
from getting trapped in that local minimizer, by moving
stochastically around it. This behavior enables the particles to
continue searching in potentially better areas of the search
space, i.e.,  closer to the global minimizer. Thus, instead of
moving directly towards the best positions, the particles will
oscillate around them. The size, $N$, of the swarm, as well as the
size of the neighborhood in the local variant of PSO can be
selected arbitrarily, although it is a common belief in
evolutionary algorithms that a population size equal from $2$ to
$10$ times the dimension of the problem at hand is a good initial
guess \citep{SP97}. Moreover, the neighborhood's size shall be
problem--dependent. In simple problems, larger neighborhoods
result in faster convergence without loss of the algorithm's
efficiency, while, in complicate problems with a plethora of local
minima, smaller neighborhoods are considered a better starting
choice.


\subsection{Detecting further periodic orbits through deflection}
\label{def}

By applying the PSO method we are able to detect one, in general
arbitrary, minimizer of the objective function. However, in our
case, several minimizers of the objective function are required,
since, in general, there exist many $p$--periodic orbits in the
search space, $S$. Restarting the PSO algorithm does not guarantee
the detection of a different minimizer. In such cases, the {\em
deflection\/} technique can be used. This technique consists of a
transformation of the objective function, $f(X)$, once a
minimizer, $X_i^*$, $i=1,\ldots,n_{\min}$, has been detected
(Magoulas, Plagianakos \& Vrahatis 1997; Parsopoulos \& Vrahatis
2002, 2004),
\begin{equation}
F(X) = \prod_{i=1}^{n_{\min}}T_i(X; X_i^*,\lambda_i)^{-1} f(X),
\label{eq:Defl1}
\end{equation}
with,
\begin{equation}
T_i(X; X_i^*,\lambda_i) = \tanh ( \lambda_i \| X - X_i^* \| ),
\label{eq:Defl2}
\end{equation}
where $\lambda_i$, $i=1,\ldots,{n_{\min}}$, are nonnegative
relaxation parameters, and $n_{\min}$ is the number of the
detected minimizers. The transformed function has exactly the same
minimizers with the original $f(X)$, with the exception of
$X_i^*$. Alternative configurations of the parameter $\lambda$
result in different shapes of the transformed function. For larger
values of $\lambda$ the impact of the deflection technique on the
objective function is relatively mild. On the other hand, using $0
< \lambda < 1$ results in a function $F(X)$ with considerably
larger function values in the neighborhood of the deflected
minimizer.

\begin{figure*}
\begin{center}
\includegraphics[scale=0.3]{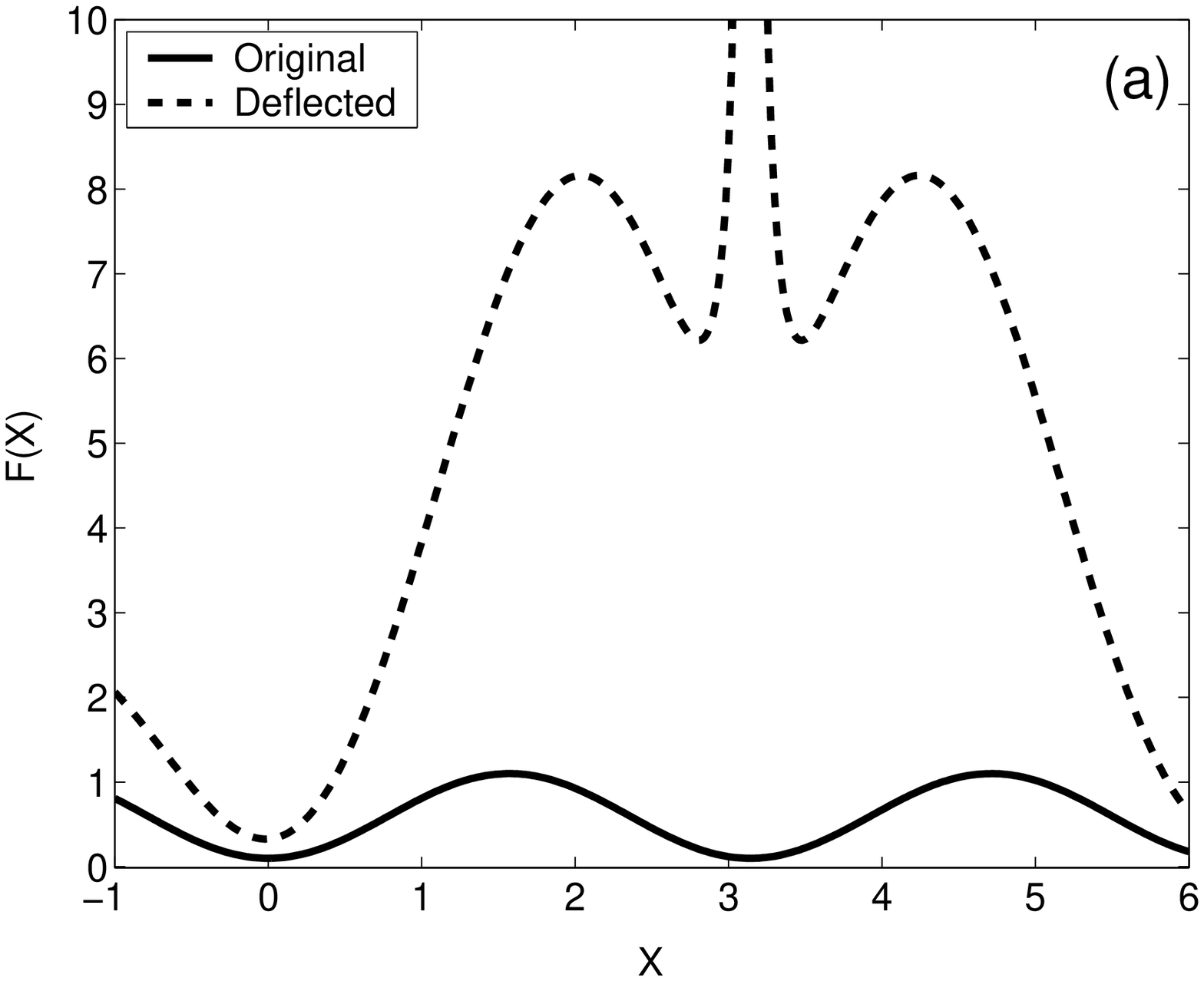} \,
\includegraphics[scale=0.3]{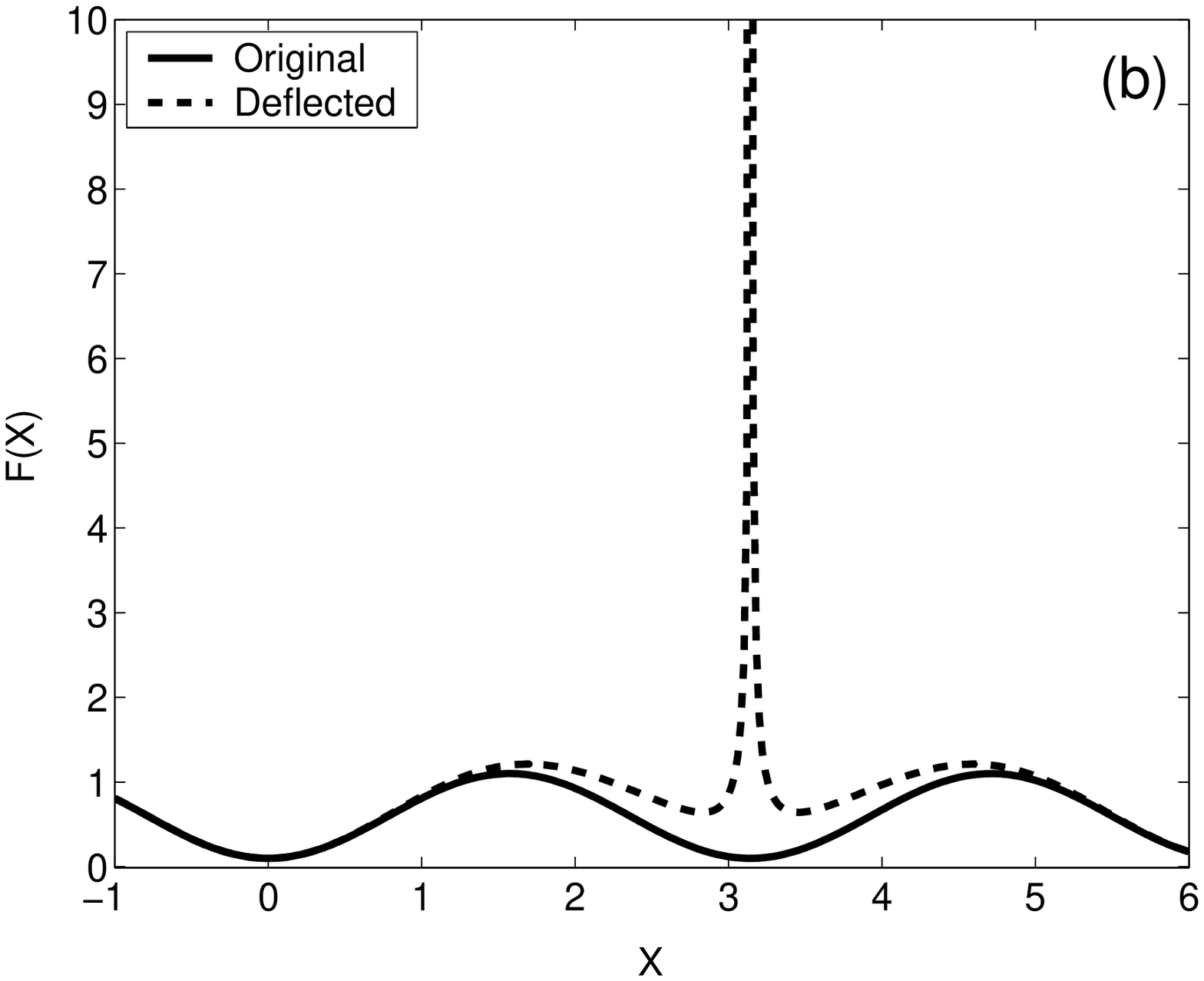} \,
\includegraphics[scale=0.3]{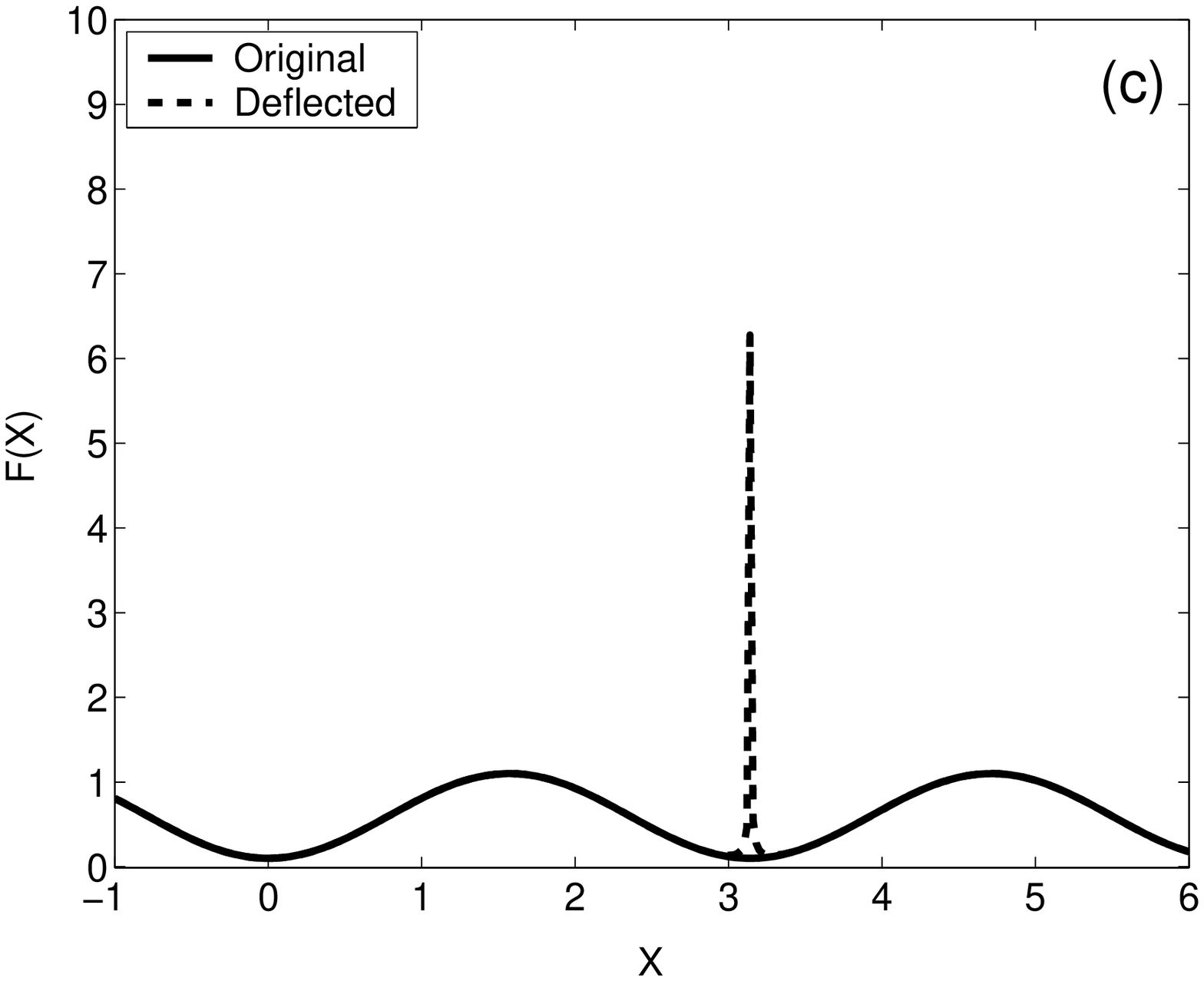}\\
\caption{The effect of the deflection procedure on the function
$f(x) = \sin^2(x) + 0.1$, at the point $x^* = \pi$, for $\lambda =
0.1$ (a), $\lambda = 1$ (b), and $\lambda = 10$ (c). }
\label{fig:PlotDef}
\end{center}
\end{figure*}

A point to notice is that the deflection technique should not be
used on its own on a function  whose global minimum is zero, as is
the case for the function $f(X)$ of Eq.~(\ref{eq:function}). The
reason is that the transformed function, $F(X)$, of
Eq.~(\ref{eq:Defl1}), will also have zero value at the deflected
global minimizer, since $f(X)$ will be equal to zero at such
points. This problem can be easily alleviated by considering the
function $\hat{f}(X) = f(X) + c$, where $c > 0$ is a constant,
instead of $f(X)$. The function $\hat{f}(X)$ possesses all the
information regarding the minimizers of $f(X)$, but the global
minimum is increased from zero to $c$ \citep{PV02b}. The value of
$c$ does not affect the performance of the algorithm and, thus, if
there is no information regarding the global minimum of $f(X)$, it
can be selected arbitrarily large. The effect of the deflection
procedure on the function $f(x) = \sin^2(x) + 0.1$, at the point
$x^* = \pi$, is illustrated in Fig.~\ref{fig:PlotDef}. In our
case, we used the  parameter values $\lambda=10^4$ and $c=0.1$ for
all the computed periodic orbits.

When the initial conditions of a $p$--periodic orbit are
determined, the rest $(p-1)$ intersections of the orbit with the
PSS (which can also be considered as initial conditions for the
same orbit) are obtained through $(p-1)$ subsequent iterations of
$\Phi(X)$ of Eq.~(\ref{eq:Phi}). We determine the stability of the
periodic orbit using established techniques \citep[see for
example][]{B69,H75,pf84,CM85,S01}. The periodic orbit can either
be stable (S) or unstable. We note that in 3D Hamiltonian systems
we can have three different types of instability, namely simple
unstable (U), double unstable (DU) and complex unstable
($\Delta$). The repetitive use of the deflection technique allows
us to find several periodic orbits of a certain period, $p$. The
proposed algorithm for the detection of periodic orbits is
described in pseudocode in Table~\ref{tab:propalgo} \citep{PV02b}.
\begin{table*}
\caption{The proposed algorithm. Note that $K$  denotes the
maximum number of the periodic orbits that can be obtained in one
run of the algorithm.}
\begin{tabular}{ll}
\hline
Input:  & Hamiltonian $H$, period $p$, maximum number of deflections $K$.\\
\hline
Step 1  & \textbf{Set} the stopping flag, $\textsl{SF} \gets 0$, and the counter, $k \gets 0$.\\
Step 2  & \textbf{While} $(\textsl{SF}=0)$\/ \textbf{Do}\\
        & \quad \quad \textbf{Apply} PSO \\
Step 3  & \quad \quad \textbf{If} (PSO detected a periodic orbit, $X_1$) \textbf{Then}\\
        & \quad \quad \quad \quad \textbf{Compute} all points $X_2,\ldots,X_p$, of the periodic orbit, \\
        & \quad \quad \quad \quad by iterating function $\Phi$
  (Eq.~(\ref{eq:Phi})) on $X_1$. \\
Step 4  & \quad \quad \quad \quad \textbf{If} $(k < K)$ \textbf{Then}\\
        & \quad \quad \quad \quad \quad \quad \textbf{Apply} Deflection on $X_1,\ldots,X_p$, and set\\
        & \quad \quad \quad \quad \quad \quad the counter $k \gets k+1$.\\
        & \quad \quad \quad \quad \textbf{Else}\\
        & \quad \quad \quad \quad \quad \quad \textbf{Set} $\textsl{SF} \gets 1$\\
        & \quad \quad \quad \quad \textbf{End If}\\
        & \quad \quad \textbf{Else}\\
        & \quad \quad \quad \quad \textbf{Write} ``No further periodic orbit was detected''\\
        & \quad \quad \quad \quad \textbf{Set} $\textsl{SF} \gets 1$\\
        & \quad \quad \textbf{End If}\\
        & \textbf{End While}\\
Step 5  & \textbf{Report} all detected periodic orbits.\\
\hline
\end{tabular} \label{tab:propalgo}
\end{table*}
\section{Numerical results}
\label{res}

Applying the scheme described in Sec.~\ref{alg}, we were able to
find, apart from the known 1--periodic orbits of Hamiltonian
(\ref{eq:Ham}) \citep{SPA02a,PSA03b}, many new 2D and 3D orbits of
period 1. We also followed these orbits as the energy change,
registering simultaneously their stability transitions and
locating  their bifurcations.

To investigate the performance of the algorithm, we searched for
periodic orbits of multiplicity 1, at the region between the
radial 4:1 and corotation resonances. In order to locate these
orbits we used as search space, $S$,  a subspace of the PSS, where
the values of the initial conditions vary in some suitable
intervals. Instead of letting all four variables, $x$, $z$,
$\dot{x}$, $\dot{z}$, to  vary simultaneously, which means that
the search space will be 4--dimensional, we adopted a more
efficient scheme, described below.

First we located the planar (2D) orbits of the system on the
equatorial plain that exist in the 4-dimensional search space $S$.
This was done in two phases. In the first phase, we located the 2D
orbits starting perpendicular to the $y=0$ axis, having initial
conditions of the form $(x, 0,0,0)$. This means that the actual
search space in this phase was 1--dimensional, as only the $x$
variable of the initial condition changed. Confining our search in
an 1--dimensional search space, also allowed us  to  decrease the
size $N$ of the swarm to $N=5$, using a smaller radius, $r=1$, of
the neighborhood of every particle. All  periodic orbits traced in
this phase were registered and the corresponding initial
conditions were transformed into maximizers of the objective
function $f(X)$ through the deflection technique. The search for
2D periodic orbits was completed in a second phase by considering
orbits starting not perpendicular to the $x$ axis, having initial
conditions of the form $(x, 0,\dot{x},0)$. In this stage the
actual search space was 2--dimensional, while the swarm consisted
of $N=10$ particles and the neighborhood radius was set to $r=2$.
As the periodic orbits with $\dot{x}=0$ had already been found and
prevented from been retraced  in the previous stage of the
application of  PSO, we located planar orbits with $\dot{x} \neq
0$. Again, these periodic orbits were registered and prevented
from been detected again, through deflection.

After the detection of the 2D periodic orbits of the system on the
equatorial plane, we searched for purely 3D periodic orbits. Again
the search was performed in different phases, keeping the
dimensionality of the search space as low as possible. Since many
3D periodic orbits found by \citet{SPA02a} have initial conditions
of the form $(x, z,0,0)$ and $(x, 0,0, \dot{z})$, we  first
concentrated our tries to detect periodic orbits of this form. For
such orbits, the actual search space of the PSO method is
2--dimensional, since  two of the four variables are set equal to
zero, so we used again $N=10$ and $r=2$. We note that orbits with
$z=0$ and $\dot{z}=0$ had already been found and, thus, in these
phases we detected actual 3D orbits. Instead of continuing our
search by letting only one variable be equal to zero and having
all the possible 3--dimensional search spaces, we preferred to
search for orbits with initial conditions of the form $(x,
z,\dot{x},\dot{z})$, because the physically most important
periodic orbits had already been found. Therefore, the complete
search for orbits with initial conditions of the form $(x,
z,\dot{x},\dot{z})$ was actually performed, using $N=20$ and
$r=3$.

The scheme of applying  PSO in several phases is more efficient
than the immediate search for orbits with initial conditions of
the form $(x, z,\dot{x},\dot{z})$, because in every phase we were
confined in search spaces of dimensionality lower than 4, which is
a computationally easier task. Additionally, we  traced the
periodic orbits in a physically meaningful way for the particular
system, as we found first the planar 2D orbits and then the
spatial 3D orbits.

\begin{figure*}
\begin{center}
\includegraphics[scale=0.43]{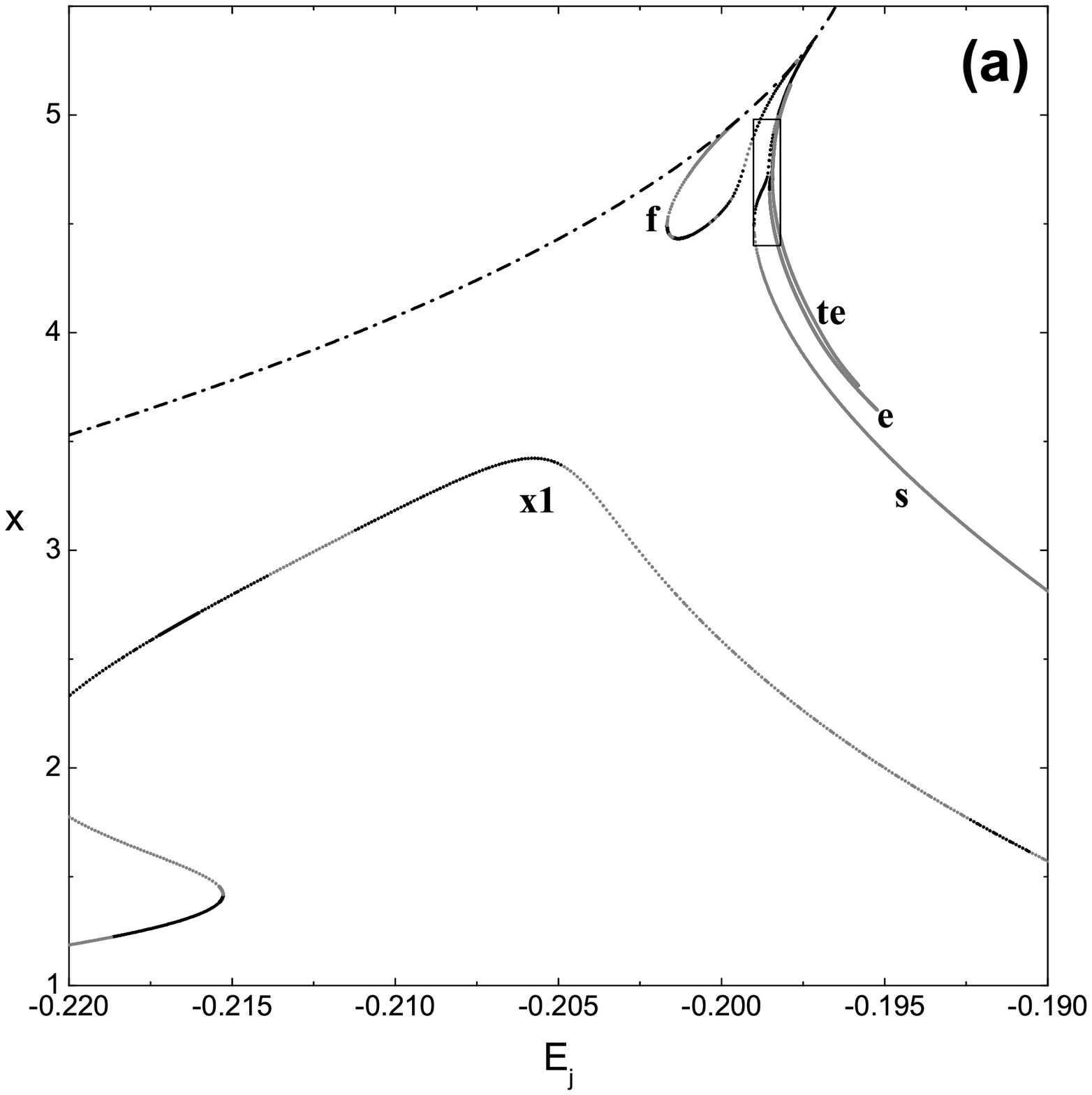} \,
\includegraphics[scale=0.43]{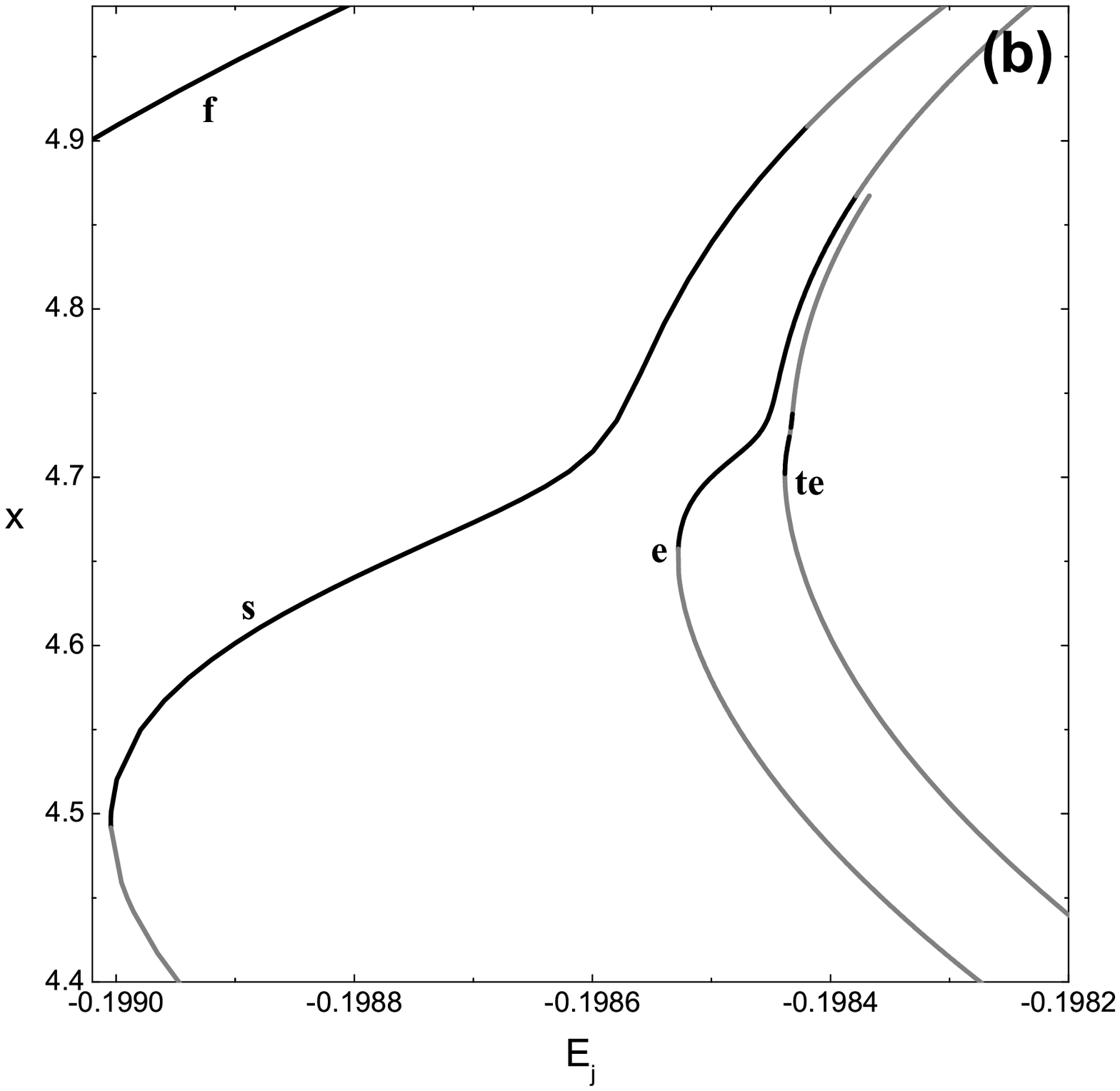}\\
\caption{(a) Characteristic diagram of families {\tt x1}, {\tt f},
{\tt s}, {\tt e}, {\tt te}. The dash-dotted curve is the section
of the zero velocity surface with the $(E_j,x)$ plane. In (b) we
give the enlargement of the region included in the small
rectangular in the upper right side of (a). } \label{fig:char}
\end{center}
\end{figure*}

The main periodic orbits that we detected by using the PSO method
are presented in the so--called `characteristic' diagram
\citep[for definition see][Section 2.4.3]{C02} shown in Fig.
\ref{fig:char}.
The diagram gives the $x$ coordinate of the initial condition of
periodic orbits as a function of the energy $E_j$. Since we
present in Fig. \ref{fig:char} only planar periodic orbits (i.e.,
$z=\dot{z}=0$), starting perpendicular to the $x$ axis (i.e.,
$\dot{x}=0$) with $y=0$ and $\dot{y}>0$, every point defines
completely the initial condition of the orbit. Orbits on the
$(x,y)$ plane with $\dot{x}\neq 0$ and orbits not lying on the
$(x,y)$ plane are not represented on the characteristic diagram of
Fig. \ref{fig:char}. We also note that black curves correspond to
stable periodic orbits, while gray curves indicate unstable
periodic orbits without discriminating between their different
instability types. In Fig. \ref{fig:char},
 we plot parts of the characteristic curves of
the main 1--periodic family of the system, the so--called {\tt x1}
family \citep{CP80}, and of the known \citep{PSA03b} 2D {\tt f}
and {\tt s} families, as well as the curves of two new families
that were found using the PSO method. The characteristic curves of
these two families are very close to each other and to the
characteristic curve of family {\tt s}. So, in order to clearly
see them we enlarge in Fig.~\ref{fig:char}(b) the region enclosed
in the small rectangular of Fig. \ref{fig:char}(a). Also,
 we can see in Fig.~\ref{fig:char}(b) how close to each other are the
various families of periodic orbits in this region of the
characteristic diagram. This is the main difficulty that hinders
other classical methods based on Newtonian techniques from
locating periodic orbits in this region.

The two new families obtained through the proposed approach, exist
for energies greater than a minimum energy value, similarly to
families {\tt f} and {\tt s}. Since close to these minimum energy
values the morphology of these families is influenced by the 8:1
and 10:1 resonances, we named these families as {\tt e} and {\tt
te} families, respectively.

\begin{figure*}
\begin{center}
\includegraphics[scale=1.6]{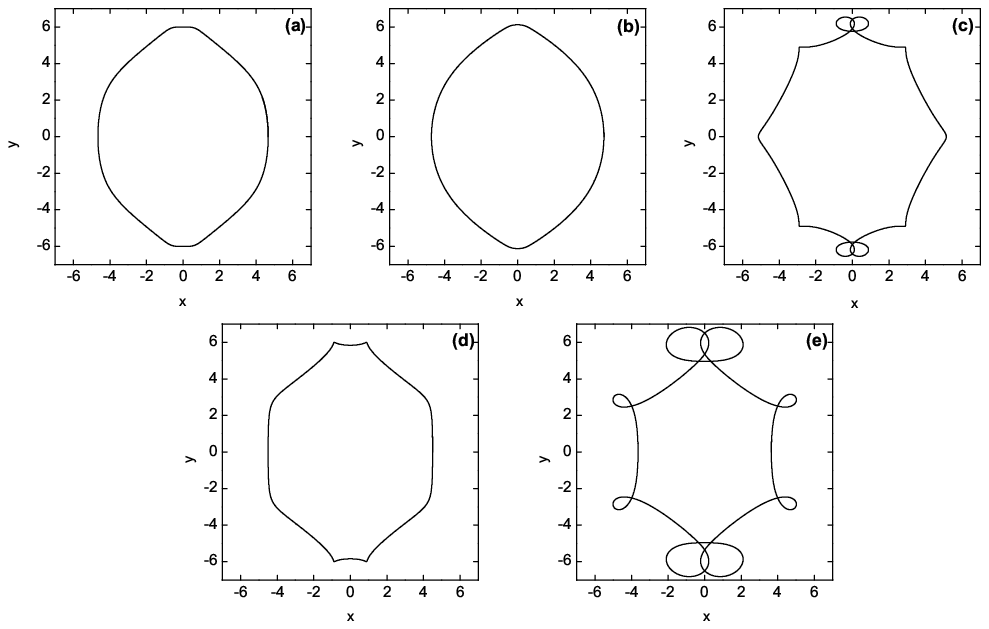}
\caption{Orbits of the 2D {\tt e} family. From (a) to (b) and then
to (c) we see the evolution of the orbital morphology along the
upper branch of the characteristic of the {\tt e} family as energy
increases, while from (d) to (e) we see the evolution along the
lower branch of the characteristic plotted in Fig. \ref{fig:char}.
Orbits (a) and (b) are stable while orbits (c), (d) and (e) are
simple unstable. We remind that the bar major axis lies along the
$y$ axis and the units are in Kpc. } \label{fig:e}
\end{center}
\end{figure*}

The {\tt e} family exists for $E_j>-0.19853$. In Fig.
\ref{fig:char}(b), we see that the characteristic curve of the
{\tt e} family has two branches with stable orbits existing at the
upper branch. In Fig. \ref{fig:e}, we plot orbits of the {\tt e}
family for increasing values of the energy belonging to the upper
branch (orbits (a), (b) and (c)), as well as to the lower branch
(orbits (d) and (e)) of the characteristic.
Along the upper branch, we observe a smooth transition from a
basically 8:1 morphology (orbit (a)) to a 10:1 morphology (orbit
(c)) through an oval--shaped configuration (orbit (b)). We note
that, as the energy increases, the orbits  develop `corners' along
the minor axis of the bar ($x$ axis). An analogous transition from
a 6:1 to an 8:1 morphology appears at the upper branch of the
characteristic of the {\tt s} family, as can be seen by comparing
Figs. \ref{fig:e}(a), (b) and (c) with Figs. 6(a), (b) and (c) of
\citet{PSA03b}. On the other hand, the morphology of the {\tt e}
family at the lower branch of its characteristic is always
influenced by the 8:1 resonance (orbits (d) and (e)), and, as
energy increases, the orbits develop loops (orbit (e)). We also
note that the unstable orbits of the lower branch of the
characteristic have almost straight segments parallel to the bar
major axis ($y$ axis). Similar morphological evolution was
observed also for the {\tt s} family as can be seen by comparing
Figs. \ref{fig:e}(d) and (e) with Figs. 6(e) and (f) of
\citet{PSA03b}.

\begin{figure*}
\begin{center}
\includegraphics[scale=1.7]{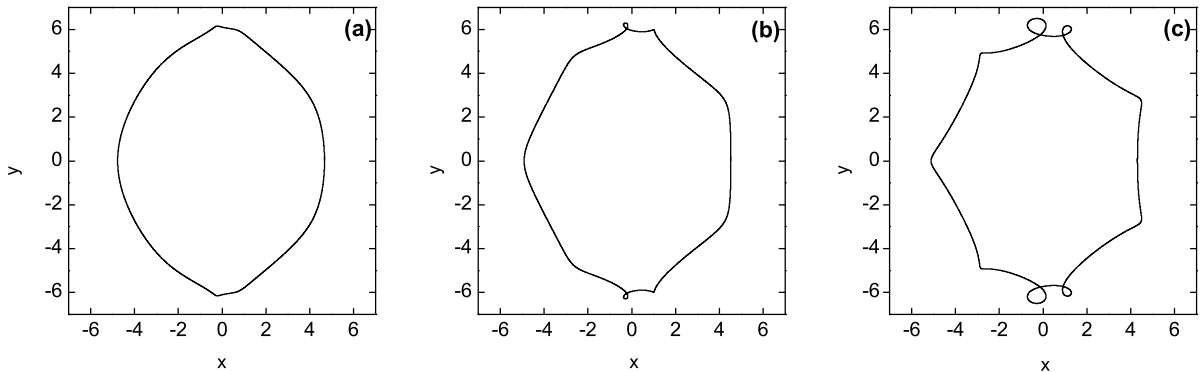}
\caption{Orbits of the 2D family {\tt er1}. From (a) to (b) and
then to (c) we see the orbital evolution of the family as the
energy increases, moving away from its bifurcation point from
family {\tt e}. Orbits (a) and (b) are stable, while orbit (c) is
simple unstable. } \label{fig:er1}
\end{center}
\end{figure*}

A new 2D family is born by bifurcation from the {\tt e} family at
$E_j \approx -0.19847$. We call it {\tt er1} following  the
nomenclature in \citet{SPA02a}. The {\tt er1} family has a 9:1
morphology (Fig.~\ref{fig:er1}), in analogy to the {\tt sr1}
family (Fig.~6(d) of \citet{PSA03b}), which bifurcates from the
{\tt s} family and it is influenced by the 7:1 resonance.

\begin{figure*}
\begin{center}
\includegraphics[scale=1.7]{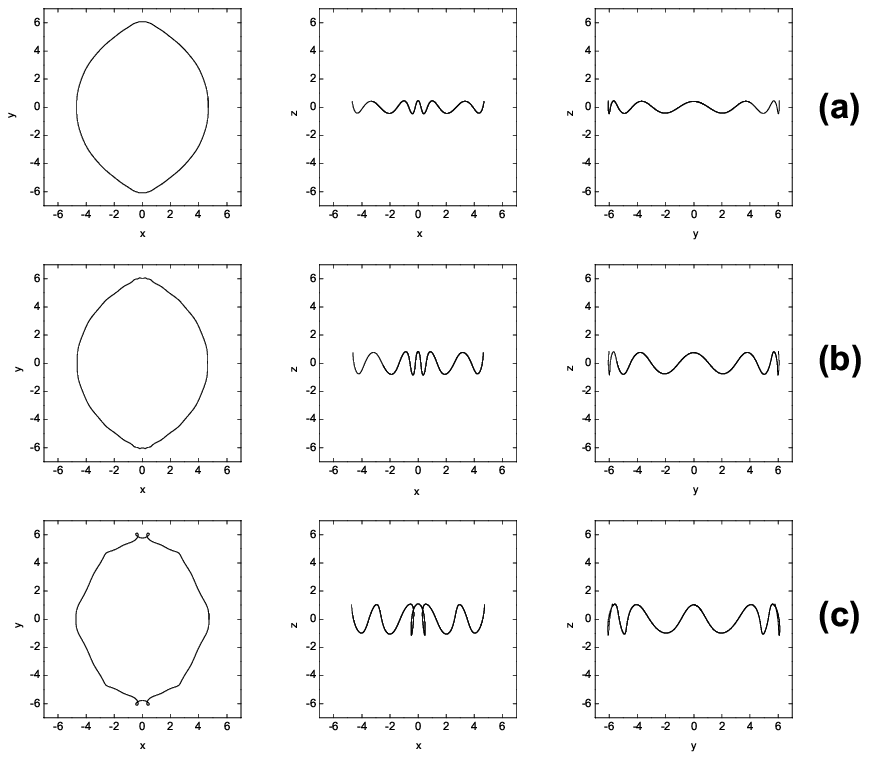}
\caption{3D stable orbits of the {\tt ev1} family. For every orbit
we plot its projection on the $(x,y)$, $(x,z)$ and $(y,z)$ planes.
From (a) to (b) and then to (c) the energy increases moving away
from the bifurcation point of {\tt ev1} from the {\tt e} family.}
\label{fig:ev1}
\end{center}
\end{figure*}

In Fig.~\ref{fig:ev1}, we see the morphological evolution of the
{\tt ev1} 3D family, which bifurcates from the {\tt e} family at
$E_j \approx -0.19849$.
We note that the projections of the {\tt ev1} orbits on the
$(x,y)$ plane are similar to the planar 2D orbits of the {\tt e}
family for the same energies. So the {\tt ev1} orbits appear as
oval-shaped and evolve to a 10:1 morphology.

\begin{figure*}
\begin{center}
\includegraphics[scale=1.6]{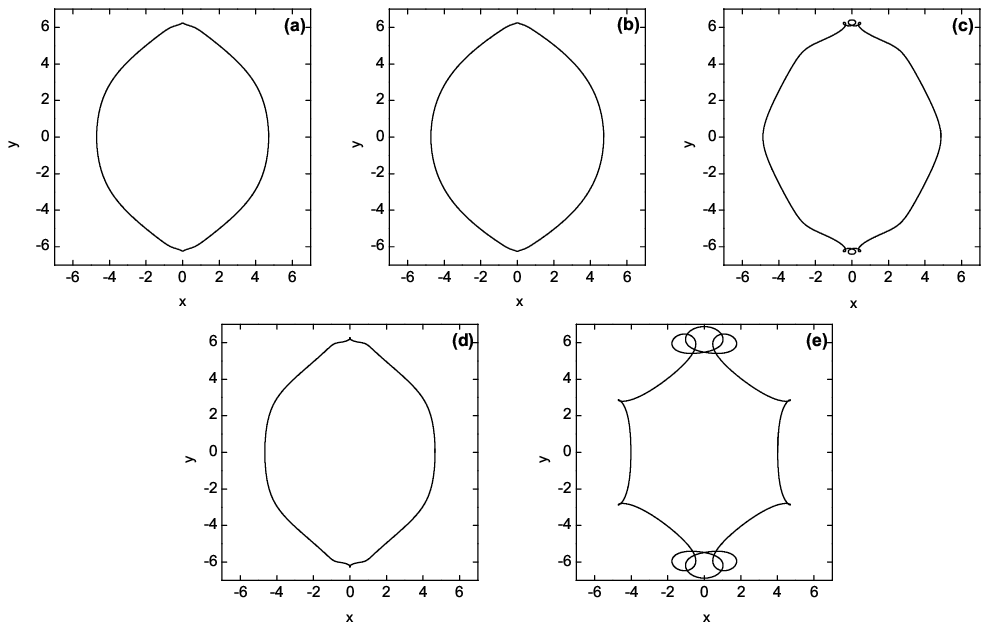}
\caption{Orbits of the 2D {\tt te} family. From (a) to (b) and
then to (c) we see the evolution of the orbital morphology along
the upper branch of the characteristic of the {\tt te} family as
energy increases, while from (d) to (e) we see the evolution along
the lower branch of the characteristic depicted in
Fig.~\ref{fig:char}. Orbits (a) and (b) are stable, orbits (c) and
(e) are double unstable and orbit (d) is simple unstable. }
\label{fig:te}
\end{center}
\end{figure*}

\begin{figure*}
\begin{center}
\includegraphics[scale=1.7]{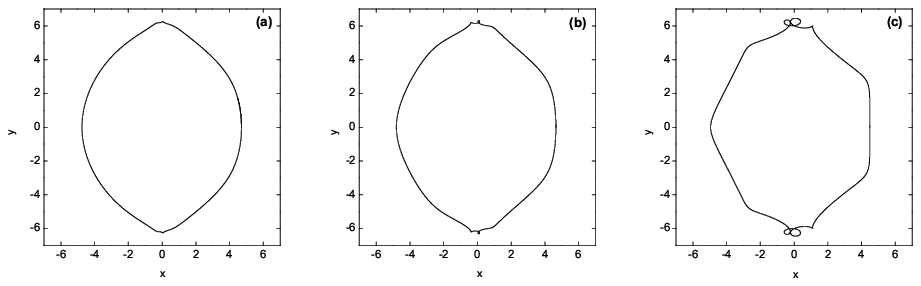}
\caption{Orbits of the 2D family {\tt ter1}. From (a) to (b) and
then to (c) we see the orbital evolution of the family as the
energy increases, moving away from its bifurcation point from
family {\tt te}. Orbit (a) is stable, while orbits (b) and (c) are
simple unstable. } \label{fig:ter1}
\end{center}
\end{figure*}

\begin{figure*}
\begin{center}
\includegraphics[scale=1.7]{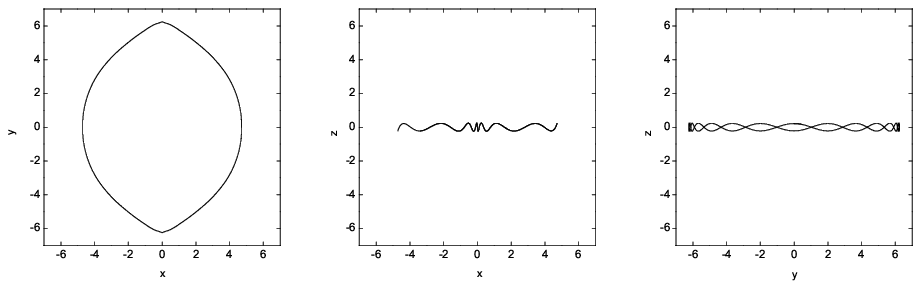}
\caption{A 3D stable orbit of the {\tt tev1} family.}
\label{fig:tev1}
\end{center}
\end{figure*}

The {\tt te} family exists for $E_j>-0.198438$, and it has similar
evolution to the {\tt e} family. At the upper branch of its
characteristic curve, there exist stable representatives of the
family (orbits (a) and (b) of Fig.~\ref{fig:te}), while, at the
lower branch, all orbits are unstable (orbits (d) and (e) of Fig.
\ref{fig:te}).
Family {\tt te} has a 10:1 morphology (Fig.~\ref{fig:te}(a)),
which evolves to a morphology influenced by the 12:1 resonance
(Fig. \ref{fig:te}(c)). The orbits belonging to the upper branch
of the characteristic curve develop `corners' along the $x$ axis
(Fig. \ref{fig:te}(c)), while the orbits of the lower branch have
segments parallel to the $y$ axis (Figs. \ref{fig:te}(d) and (e)).
The 2D family {\tt ter1} that bifurcates from family {\tt te}  at
$E_j \approx -0.198434$ is influenced by the 11:1 resonance (Fig.
\ref{fig:ter1}),
while the 3D family {\tt tev1}, which bifurcates from  family te
at $E_j \approx -0.198435$, has an oval--shaped projection on the
$(x,y)$ plane (Fig. \ref{fig:tev1}).

\section{Discussion and conclusions}
\label{con}

In this paper, we proposed and applied an algorithm for locating
periodic orbits in a 3D galactic potential. Our numerical scheme
is based on the transformation of a root--finding problem, i.\,e.,
solving Eq.\,(\ref{eq:eq}), into a problem of detecting  the
global minimizers of function $f$ (Eq.\,\ref{eq:function}). The
detection of the minimizers is performed by the PSO method. This
transformation also enables us to use a deflection technique, so
that many periodic orbits are traced in one run of the algorithm.
By using  the proposed algorithm we were able to trace 2D and 3D
periodic orbits close to the corotation region in the barred
galaxy model described in Sec.\,\ref{pot}. Our numerical results
justify the usefulness of the new scheme, as well as its
advantages, which springs from its ability to locate periodic
orbits in regions where many periodic orbits coexist very close to
each other, even when many of them are unstable. In addition, even
if a periodic orbit is found, Newton--like methods cannot
guarantee that repeating the search with a different initial guess
will result in  a new periodic orbit. The proposed scheme
alleviates this problem.

In order to demonstrate the effectiveness of the proposed scheme,
we discuss now in detail its application on the simplest case that
we faced in the present paper. This case was the location of
planar, 2D, orbits of period $p=1$, starting perpendicular to the
$y=0$ axis, having initial conditions of the form $(x, 0, 0, 0)$.
Restricting the search for periodic orbits close to the corotation
we set $E_j=0.1984$ and let $x$ varying in the region $3.5
\leqslant x \leqslant 5.5$. In this case the search space of the
PSO method was 1--dimensional and so we used a swarm of $N=5$
particles, while the neighborhood of every particle had radius
$r=1$. Every iteration of the PSO method performed the evaluation
of the objective function $f(X)$ (Eq.\,\ref{eq:function}) with
$p=1$, for $N=5$ times. This means that we computed the first
intersection with the PSS of $N=5$ orbits or, in other words, we
performed $N=5$ evaluations of $\Phi(X)$ (Eq.\,\ref{eq:Phi}),
which  is the most time consuming computation in every iteration
of the PSO method. In the particular application we asked the
tracing of up to 15 periodic orbits in one run of the algorithm
and we succeeded in locating all the planar 2D orbits of families
{\tt e}, {\tt te}, {\tt er1} and {\tt ter1} existing in the search
space, as well as, several banana--like orbits (not discussed in
the paper) and orbits belonging to the {\tt s} family. All the
periodic orbits were located with accuracy of at least $10^{-10}$.
The whole run performed about 1350 iterations of the PSO method
(which means about 90 iterations per periodic orbit), involving a
total of about 6750 evaluations of $\Phi(X)$. A main advantage of
our approach is that it does not need good initial guesses close
to the real initial conditions of the periodic orbits it traces.
As an example we refer to the periodic orbit belonging to the {\tt
ter1} family that was found by the above mentioned run. This orbit
is simple unstable and the usual Newton iterative method for
finding it, converges to its initial condition $x_0$ for initial
guesses of $x$ satisfying $|x-x_0|\lesssim 7 \cdot 10^{-5}$, after
about 10 successive iterations. We note that in the general case
of a 4--dimensional PSS every iteration of the Newton scheme
involves the computation of the Jacobian matrix of the
corresponding set of equations, which requires 4 evaluations of
$\Phi(X)$ \citep[see for example][]{pf93}. Thus, in principle, in
order to be able to locate all the 2D periodic orbits in the
interval $x \in [3.5, 5.5]$ found by the PSO method, one should
compute all the orbits with initial conditions on a grid of this
interval whose grid step should at least be equal to $7 \cdot
10^{-5}$. These procedure requires about 28500 evaluations of
$\Phi(X)$, not taking into account the successive iterations of
the Newton method to converge to the actual initial condition with
the desired accuracy. We note that in the case of the {\tt ter1}
periodic orbit this convergence requires about 10 evaluations of
$\Phi(X)$. So it is evident that the computational effort needed
by our numerical scheme is significantly lower with respect to the
Newton iterative method even in the simple case of the
1--dimensional search space.

As we pointed out, the main problem that Newton--like techniques
face in cases where many periodic orbits coexist very close to
each other, is that they need a very good initial guess of the
position of the periodic orbit in order to find it. This
difficulty of the Newton iterative schemes is well known and some
efforts to improve their efficiency have already been done. As an
example, we refer to the paper of \cite{pf93} where the authors
evaluate the initial conditions in every successive iteration of
the Newton scheme by an appropriate least squares technique. We
emphasize that our approach is completely different as we do not
try to find the roots of the system of Eq.\,(\ref{eq:eq}) but we
transform this problem into a minimization one. We also note that
our scheme uses only the values of function $\Phi$ (Eq.\,
\ref{eq:Phi}) and not its derivatives like Newton iterative
techniques do. This makes our computation easier and more
accurate, since the usual method to evaluate the derivatives of
function $\Phi$, which is not known in a close analytical form, is
to approximate them by finite differences.

Summarizing the main advantages of the proposed algorithm with
respect to Newton--like methods we could mention that our method
is faster, it is simple and can be implemented easily, it works
using function values solely, and also it has the ability to
locate many periodic orbits per run.

The orbital behavior of the {\tt e} and {\tt te} families and
their bifurcations, which were detected by the proposed scheme, as
well as the behavior of the {\tt s} family \citep{PSA03b}, helps
us to establish the general behavior of the families existing
beyond the radial 4:1 resonance gap in our model. The morphology
of the basic 2D families are influenced by two successive even
resonances: the {\tt s} family is influenced by the 6:1 and 8:1
resonances, the {\tt e} family is influenced by the 8:1 and 10:1
resonances and the {\tt te} family by the 10:1 and 12:1
resonances. From these families, 2D families influenced by the
in-between odd resonances bifurcate: {\tt sr1} is influenced by
the 7:1 resonance, {\tt er1} by the 9:1 resonance and {\tt ter1}
by the 11:1 resonance. In addition, the main 3D bifurcations of
the {\tt s}, {\tt e} and {\tt te} families ({\tt sv1}, {\tt ev1},
{\tt tev1} families respectively) have, in general, projections on
the $(x,y)$ plane similar to the main families. This is in
accordance with the observed frequency of the various inner rings
morphology \citep{B95}, as proposed by \citet{PSA03b}.

\section*{Acknowledgments}
We appreciate useful comments from the referee, which helped us
improve the clarity of the paper. We also thank Prof. G.
Contopoulos for fruitful discussions on the subject. Ch. Skokos
was supported by the Research Committee of the Academy of Athens,
the `Karatheodory' post--doctoral fellowship No 2794 of the
University of Patras, the Greek State Scholarships Foundation
(IKY) and the EMPEIRIKEION Foundation.

\end{document}